\documentclass[lettersize,journal,twocolumn]{IEEEtran}
\usepackage{graphicx}
\graphicspath{{Figures/}}
\usepackage{amsfonts}
\usepackage{amssymb}
\usepackage{amsmath}
\usepackage{color}
\usepackage{wasysym}
\usepackage{cite}
\usepackage{hyperref}
\usepackage{bbold}
\usepackage{algpseudocode}
\usepackage{placeins}
\usepackage{makecell}
\usepackage{braket}

\usepackage{mathtools}

\definecolor{lightblue}{RGB}{136,163,209}

\usepackage{algorithm}
\usepackage{algpseudocode}

\usepackage{authblk}

\begin{document}

\author[1]{Matteo Vandelli}
\affil[1]{Quantum Computing Solutions, Leonardo S.p.A., Via R. Pieragostini 80, Genova, 16151, Italy}
\author[1]{Francesco Ferrari}
\author[1,2]{Daniele Dragoni}
\affil[2]{Leonardo Hypercomputing Continuum, Leonardo S.p.A., Via R. Pieragostini 80, Genova, 16151, Italy}

\title{Learning Grover mixers for constrained optimization}

\maketitle

\begin{abstract}
Grover-Mixer QAOA (GM-QAOA) is a promising quantum framework for constrained combinatorial optimization. However, its practical implementation requires the efficient preparation of a uniform superposition over feasible states. In this work, we propose a two-stage variational approach. First, a parameterized quantum circuit is trained to prepare this superposition, using an entropy-based cost function that favors a uniform distribution over feasible bitstrings. The trained circuit is then embedded in GM-QAOA, where it provides both the initial state and, together with its inverse, yields the Grover mixer. The resulting method is evaluated on two representative benchmark problems: the Maximum Weight Independent Set (MWIS) problem and the Multidimensional Knapsack Problem (MKP). While the results demonstrate consistent improvements over standard QAOA for MWIS, only marginal gains are observed for MKP, with the advantage diminishing as the problem size increases. The reduced performance on MKP reflects the growing difficulty of training the variational mixer, likely due to increasingly challenging optimization landscapes. Overall, these findings highlight both the potential of variational Grover mixers and the trade-off between their performance gains and the additional resources required for their training.
\end{abstract}

\section{Introduction}

\begin{equation}
\begin{aligned}
    \min_{x \in \{0,1\}^n}   f(x), \    \text{subject to} \  \boldsymbol{g}(x) \leq 0  
\end{aligned}
\label{eq:const_opt}
\end{equation}
\noindent is the generic formulation of a constrained binary optimization problem, in which an objective function $f(x)$ is minimized over a binary decision vector $x$, subject to $m$ constraints defined by the functions $\boldsymbol{g}(x)=(g_1(x), \dots, g_m(x))$. Problems of this kind arise in numerous industrial decision-making applications~\cite{petropoulos2024or}. Many of these problems are NP-hard, making their efficient solution a significant computational challenge. Classical exact methods, such as branch-and-bound~\cite{land1960bb, papadimitriou1982combinatorial, nemhauser1988integer}, guarantee optimality but require exponential time in the worst case, whereas metaheuristics~\cite{BOUSSAID201382}, such as local search and evolutionary algorithms, can efficiently explore large solution spaces, but provide no optimality guarantees.

The binary structure of Eq.~\eqref{eq:const_opt} makes these problems direct candidates for quantum optimization algorithms~\cite{abbas2024review}: by identifying each variable with a qubit, the objective can be expressed as an Ising-like Hamiltonian~\cite{lucas2014ising} whose ground state encodes the optimal solution. The digitized quantum adiabatic algorithm~\cite{farhi2000quantum} and the Quantum Approximate Optimization Algorithm (QAOA)~\cite{farhi2014quantumapproximateoptimizationalgorithm}, together with its many variants~\cite{blekos2024qaoa}, are among the most widely adopted algorithms for finding such ground states. In their standard formulation, however, these algorithms are designed for unconstrained problems, and incorporating the constraints $\boldsymbol{g}$, which are ubiquitous in actual applications, remains a key bottleneck. The most common approach consists of including the constraints into the objective as penalty terms, recasting the problem as a quadratic unconstrained binary optimization (QUBO) problem~\cite{glover2022qubo}. This simplicity comes at a cost: the algorithm explores the full $2^n$-dimensional Hilbert space instead of the (usually) much smaller feasible subspace, and the penalty weights require careful tuning. Together, these effects make the optimization harder and undermine the effectiveness of QAOA~\cite{gabbassov2024lagrangian, shirai2025compressed, gabbassov2026principles}.

A better strategy is to enforce constraint satisfaction directly through the QAOA mixer, using ad hoc operators that keep the evolution within the feasible subspace, thus removing the need for penalty terms~\cite{hadfield2019qaoa}. For simple constraints, such as fixed-Hamming-weight conditions, such mixers can be built from local operators that connect feasible states differing in only a few bits~\cite{hadfield2019qaoa,PhysRevA.101.012320}. For complex, non-local constraints, constructing them becomes considerably harder, and the resulting circuits can be demanding in terms of depth~\cite{a15060202, vandelli2025constraint, awasthi2026constraint}. Moreover, even when a suitable mixer can be defined, there may be no known circuit that prepares its ground state~\cite{tsvelikhovskiy2026expressive}, which conveniently serves as the initial state.

An alternative route to enforce feasibility is taken by the Grover-mixer QAOA (GM-QAOA)~\cite{baertschi2020grover} approach: rather than engineering local constraint-preserving transitions, it builds the mixer Hamiltonian $\mathcal{H}_M$ directly from the initial state. Specifically, ${\mathcal{H}_M = |F\rangle\langle F|}$ is the projector onto the uniform superposition $|F\rangle$ of all feasible states, coupling every feasible state to every other in a single step. Mixer design and initial-state preparation thus reduce to a single task: constructing a unitary $U_F$ with ${U_F|0\rangle^{\otimes n} = |F\rangle}$. Given $U_F$, the evolution operator $e^{-i\beta |F\rangle\langle F|}$ can be implemented with a single call to $U_F$ and one to its inverse, and it confines the dynamics to the feasible subspace by construction. GM-QAOA therefore turns constraint handling into a state-preparation problem, and it is only as general as our ability to build $U_F$, which is however a non-trivial task for arbitrary constraints.

In this work, we approximate $U_F$ of GM-QAOA with parameterized quantum circuits (PQCs) trained to prepare the feasible superposition $|F\rangle$, providing a general alternative to problem-specific constructions. We test this approach on the Maximum Weight Independent Set (MWIS) problem and the Multidimensional Knapsack Problem (MKP), assessing both the feasibility of the prepared state and its impact on GM-QAOA performance. For MWIS, our approach consistently outperforms standard QAOA at comparable circuit depth, while for MKP the advantage diminishes as the problem size increases. Notably, the method bypasses slack variables, reducing the qubit count for inequality-constrained problems. At the same time, it highlights ongoing training challenges characteristic of variational quantum circuits for this application.

\section{Theory and methods}

\subsection{QUBO-based and Grover-mixer QAOA}

We consider the constrained binary optimization problem~\eqref{eq:const_opt}, with $x = (x_1, \dots, x_n) \in \{0,1\}^n$, and denote by $x_{\rm best}$ the optimal feasible solution. In QUBO-based quantum approaches, one defines an Ising-like cost Hamiltonian 
\begin{equation}
    \mathcal{H}_{\rm QUBO} = \mathcal{H}_f + \sum_{j=1}^m \lambda_j \mathcal{H}_{g_j},
    \label{eq:cost_ham}
\end{equation}
where both terms are diagonal in the computational basis: $\mathcal{H}_f|x\rangle = f(x)|x\rangle$ encodes the objective function, while each penalty term $\mathcal{H}_{g_j}$ vanishes on states satisfying the $j$th constraint (i.e., $g_j(x)\leq 0)$ and is positive otherwise. For general problems, the penalty term can require the introduction of $n_s$ slack variables in addition to the $n$ variables of the problem~\cite{glover2022qubo}. For sufficiently large weights $\lambda_j > 0$, the ground state of $\mathcal{H}_{\rm QUBO}$ is $|x_{\rm best}\rangle$. Vanilla QAOA~\cite{farhi2014quantumapproximateoptimizationalgorithm} for QUBO problem aims at approximating this ground state with the parameterized state
\begin{equation}
    |\psi_{\rm QAOA}(\boldsymbol{\beta}, \boldsymbol{\gamma})\rangle = \prod_{k=1}^{L} e^{-i \beta_k \mathcal{H}_M}\, e^{-i \gamma_k \mathcal{H}_{\rm QUBO}}\, H^{\otimes Q} |0\rangle^{\otimes Q}.
    \label{eq:qaoa}
\end{equation}
where the total number of qubits is $Q=n+n_s$. The corresponding quantum circuit first applies a layer of Hadamard gates ($H$) on all qubits, followed by $L$ layers of alternating evolution operators under $\mathcal{H}_{\rm QUBO}$ and under the transverse-field mixer Hamiltonian $\mathcal{H}_M = \sum_{i=1}^{Q} X_i$, where $X_i$ is the Pauli-$X$ operator on qubit $i$. The vectors of parameters $\boldsymbol{\beta}, \boldsymbol{\gamma} \in \mathbb{R}^L$ are optimized and the resulting state is sampled to obtain candidate solutions. While the standard loss function for the optimization of QAOA is $\langle \psi_{\rm QAOA} | \mathcal{H}_{\rm QUBO} | \psi_{\rm QAOA} \rangle$, in this work we adopt the feasibility-aware loss proposed in Ref.~\cite{ferrari2026feasibility} (introduced in Section~\ref{sec:fullcircuitopt}).

In QUBO-based approaches feasibility is thus enforced only by energetic penalties within $\mathcal{H}_{\rm QUBO}$, whose weights require careful tuning; furthermore, in practice, the probability of sampling a feasible bitstring decreases rapidly with problem size. GM-QAOA~\cite{baertschi2020grover}, on the other hand, removes the need for penalties by directly restricting the quantum state to the feasible subspace $F = \{x \in \{0,1\}^n \,|\, \boldsymbol{g}(x) \leq 0\}$. The circuit of GM-QAOA is initialized in the uniform superposition of feasible states
\begin{align}
    |F\rangle \equiv U_F|0\rangle^{\otimes n} = \frac{1}{\sqrt{|F|}} \sum_{x \in F} |x\rangle.
    \label{eq:exact_F}
\end{align}
In the following, we consider the case in which $U_F$ can be implemented using only $n$ qubits.
The final quantum state is then constructed as follows:
\begin{align}
|{\psi}_{\rm GM}(\boldsymbol{\beta}, \boldsymbol{\gamma})\rangle = \prod_{k=1}^{L} \left( U_{\rm GM}(\beta_k) \; e^{-i \gamma_k {H}_f} \right) |F\rangle.
\label{eq:gm_qaoa}
\end{align}
Compared to Eq.~\eqref{eq:qaoa}, the initial state $H^{\otimes n}|0\rangle^{\otimes n}$ is replaced by $|F\rangle$, the cost unitary contains only $\mathcal{H}_f$, and the transverse-field mixer is replaced by the Grover mixer $U_{\rm GM}$
\begin{align}
    U_{\rm GM}(\beta) &= e^{-i \beta |F \rangle \langle F|} \notag \\ &= U_F \left[\mathbb{1}_n - \left(1 - e^{-i\beta}\right) \big(|0\rangle \langle 0| \big)^{\otimes n} \right]U_F^\dagger,
    \label{eq:gm_mixer}
\end{align}
which induces transitions only between feasible states. The Grover mixer is therefore fully defined by $U_F$, the same unitary that prepares the initial state. For generic constraints, however, no efficient circuit construction for $U_F$ is known.

\subsection{Learning procedure for the Grover mixer}

In this work, instead of designing an exact circuit for $U_F$, which is a demanding task in general, we train a PQC $\overline{U}_F(\boldsymbol{\theta})$ to prepare an approximate version of $|F\rangle$ up to bitstring-dependent phases and use it within GM-QAOA. These phases can be ignored because they act as a diagonal operator, which commutes with the cost unitary and thus leaves the GM-QAOA sampling distribution unchanged~\cite{note_phases}. Here and in the following, we use barred symbols to denote approximate quantities. The approximate initial state is 
\begin{align}
    |\overline{F}\rangle = \overline{U}_F(\boldsymbol{\theta})|0\rangle^{\otimes n} = \sum_{x \in \{0, 1\}^n} c_x(\boldsymbol{\theta}) |x\rangle,
    \label{eq:approx_F}
\end{align}
whose parameters $\boldsymbol{\theta}$ are variationally optimized so that the coefficients $c_x(\boldsymbol{\theta})$ are as close as possible to a uniform superposition of feasible bitstrings, namely $p_x=|c_x|^2 = 1/|F|$ for $x \in F$ and $c_x = 0$ otherwise. As a suitable cost function for this learning step, we choose to maximize the classical entropy over feasible states only, given by
\begin{align}\label{eq:S_F}
    \mathcal{S}_F(\boldsymbol{\theta}) &= - \sum_{x \in F} p_x(\boldsymbol{\theta}) \log [p_x(\boldsymbol{\theta})].
\end{align}
Its maximum, $\mathcal{S}_F = \log|F|$, is reached when the probability is distributed uniformly over the feasible bitstrings and vanishes elsewhere, i.e., when $|\overline{F}\rangle$ coincides with $|F\rangle$ up to bitstring-dependent phases. On quantum hardware, $\mathcal{S}_F$ is evaluated by replacing $p_x$ with the empirical frequencies of the sampled bitstrings. Restricting the sum to $F$ then only requires checking $\boldsymbol{g}(x) \leq 0$ for each sample, which is classically efficient.

\subsection{PQCs for learning the Grover Mixer}

There are several possible Ans\"atze for $\overline{U}_F$. 
Here, we focus on two circuit architectures. The first is itself a QAOA circuit, so that vGM-QAOA effectively becomes a nested QAOA, with an inner circuit optimized beforehand to prepare the feasible superposition. A similar nested QAOA structure underlies the method of Ref.~\cite{zhou2026learned}, developed for hierarchical optimization problems. To avoid confusion with the outer QAOA, we refer to this inner circuit as a Hamiltonian variational ansatz (HVA)~\cite{PhysRevA.92.042303,PRXQuantum.1.020319}, the name commonly used for circuits of this type. In analogy with Eq.~\eqref{eq:qaoa}, we write
\begin{align}
   \overline{U}_F^{\rm HVA}\left(\boldsymbol{\theta}\right) = \left(\prod_{k=1}^{L_{\rm mix}} e^{-i \theta_{M, k} \mathcal{H}_M}\, e^{-i \mathcal{H}_{C}\left(\boldsymbol{\theta}_{C, k} \right)}\right)  H^{\otimes n}
   \label{eq:U_F_qaoa} 
\end{align}
with $L_{\rm mix}$ layers, where $\mathcal{H}_{\rm QUBO}$ is replaced by a Hamiltonian $\mathcal{H}_C(\boldsymbol{\theta}_{C,k})$ that encodes only the constraints. The explicit form of $\mathcal{H}_C$ depends on the specific problem and is discussed in Section~\ref{sec:results}. Its parameters, $\boldsymbol{\theta}_{C,k}$, are used to independently weight the penalty terms associated with different types of constraints. The variational parameters of layer $k$ are thus collectively denoted as $\boldsymbol{\theta}_k =(\boldsymbol{\theta}_{C,k}, \theta_{M,k})$, with a single mixer angle $\theta_{M,k}$. The Ansatz~\eqref{eq:U_F_qaoa} therefore contains $(1 + n_c)L_{\rm mix}$ parameters, where $n_c$ is the number of penalty terms in $\mathcal{H}_C$, independently of the number of qubits $n$.

The second circuit we consider is a PQC commonly used in variational quantum eigensolvers (VQE)~\cite{peruzzo2014variational, kandala2017hardware, cerezo2021variational}:
\begin{align}
    &\overline{U}_F^{\rm VQE}\left(\boldsymbol{\theta}\right) \nonumber \\ &= \left[\prod_{k=1}^{L_{\rm mix}}  \left(\prod_{i=1}^N {CX}_{i, (i \, {\rm mod} N) +1} \right) \left(\prod_{i=1}^N U_i(\boldsymbol{\theta}_{i, k})\right)\right] H^{\otimes n}.
    \label{eq:U_F_vqe}
\end{align}
Here $U_i$ represents generic single-qubit rotation acting on qubit $i$, with angles $\boldsymbol{\theta}_{i, k} \in \mathbb{R}^3$, and the CX gates entangle neighboring qubits in a ring. Circuits of this kind can be hard to train due to barren plateaus~\cite{mcclean2018barren, Ragone2024}. The Ansatz~\eqref{eq:U_F_vqe} has $3nL_{\rm mix}$ parameters, growing linearly with $n$.

\subsection{Full circuit optimization}\label{sec:fullcircuitopt}

The operator $\overline{U}_F(\boldsymbol{\theta}^*)$ is trained only once at the beginning of the workflow.
Once learned, it replaces $U_F$ in Eq.~\eqref{eq:gm_mixer}, yielding the approximate mixer $\overline{U}_{\rm GM}(\beta)$, and hence the state
\begin{align}
    |\psi_{\rm vGM}(\boldsymbol{\beta}, \boldsymbol{\gamma})\rangle = \prod_{k=1}^{L} \left( \overline{U}_{\rm GM}(\beta_k)\, e^{-i \gamma_k \mathcal{H}_f} \right) |\overline{F}\rangle.
    \label{eq:vgm_qaoa}
\end{align}
We refer to the resulting algorithm as \emph{variational} GM-QAOA (vGM-QAOA).

The remaining step is the optimization of the angle vectors $\boldsymbol{\beta}$ and $\boldsymbol{\gamma}$. To reduce the computational burden, in this work we restrict the parameters of both QUBO-based QAOA~\eqref{eq:qaoa} and vGM-QAOA~\eqref{eq:vgm_qaoa} to a linear-ramp schedule~\cite{montanez2025linear}, ${\beta_k = \beta \left(1 - \frac{k}{L}\right)}$, ${\gamma_k = \gamma \frac{k}{L}}$. This reduces the number of variational parameters from $2L$ to two, $\beta$ and $\gamma$. The latter are optimized by minimizing the average cost restricted to feasible bitstrings~\cite{ferrari2026feasibility},
\begin{align}
    \mathcal{L}_F(\beta, \gamma) = \sum_{x \in F} f(x)\, |\langle x|\psi\rangle|^2,
    \label{eq:loss_feasible}
\end{align}
where $\psi$ is $\psi_{\rm QAOA}$ for QAOA and $\psi_{\rm vGM}$ for vGM-QAOA. Since $f(x) \leq 0$ for all $x$ in both problems studied in this work, the loss function in Eq.~\eqref{eq:loss_feasible} is well defined, as it correctly favors feasible low-energy states (see Ref.~\cite{ferrari2026feasibility} for a detailed discussion of $\mathcal{L}_F$).

\section{Computational Results}\label{sec:results}

\subsection{Benchmark details}

To test our method, we benchmarked it on two constrained optimization problems of general interest, MWIS and MKP. Both of them involve inequality constraints, although the MWIS problem can be implemented in QUBO form without the use of slack variables~\cite{glover2022qubo}. To our knowledge, no exact and efficient circuit preparing the uniform superposition of independent sets is known for MWIS.

We evaluate the quality of the approximate initial state $|\overline{F}\rangle$, and thus of $\overline{U}_F$, with two metrics. The first is the entropy $\mathcal{S}_F$ of Eq.~\eqref{eq:S_F}, i.e., the training objective itself. Its maximum, $\log|F|$, is reached only when the probability is uniform over $F$ and vanishes elsewhere, so $\mathcal{S}_F$ accounts for both uniformity and feasibility but does not distinguish between them. To isolate the latter, we define the \emph{feasibility} of a generic state $|\psi\rangle$ as the total probability it assigns to feasible bitstrings,
\begin{align}
    \mathcal{F}[\psi] &= \sum_{x \in F} |\langle x|\psi\rangle|^2,
    \label{eq:fidelity}
\end{align}
and evaluate it on $|\overline{F}\rangle$. Its complement, $1 - \mathcal{F}[\overline{F}]$, quantifies the \emph{leakage}, i.e., the probability weight lost to infeasible states; a feasibility close to $1$ thus indicates negligible leakage.

While the above metrics characterize the learned initial state alone, we assess the performance of the full vGM-QAOA, and of QAOA for comparison, with three further metrics. The first is the \emph{approximation ratio}, defined here as~\cite{ferrari2026feasibility}
\begin{align}
    \alpha = \frac{\mathcal{L}_F}{f(x_{\rm best})},
\end{align}
the second is the probability $p_{\rm best}$ of sampling the optimal solution $x_{\rm best}$ from the final state, and the third is the feasibility of the final state itself, $\mathcal{F}[\psi_{\rm vGM}]$.

All quantum circuits were constructed using the IBM Qiskit framework~\cite{javadiabhari2024quantumcomputingqiskit} and emulated using the StateVector option of the Aer simulator. The classical optimization procedures were implemented using a custom implementation of fast simulated annealing (FSA)~\cite{SZU1987157}, followed by a local refinement stage based on the COBYQA optimization algorithm implemented in SciPy~\cite{2020SciPy-NMeth}. Specifically, the learning stage employed 500 iterations of FSA, followed by 100 iterations of COBYQA to refine the resulting parameter estimates. This two-stage optimization strategy combines the exploratory capabilities of simulated annealing with the local refinement capabilities of COBYQA. The whole optimization scheme is derivative-free, which is advantageous for noisy quantum circuits when executed on real quantum hardware.

For each problem configuration, we generated 16 random problem instances using NumPy~\cite{harris2020array}. These instances were used to evaluate the performance of the proposed methods across different realizations of the problem. All numerical experiments were performed on the davinci-1 proprietary computing cluster.

\subsection{Maximum Weight Independent Set}

\begin{figure}[t]
    \centering
    \includegraphics[trim={0 1.2cm 0 0}, width=\linewidth]{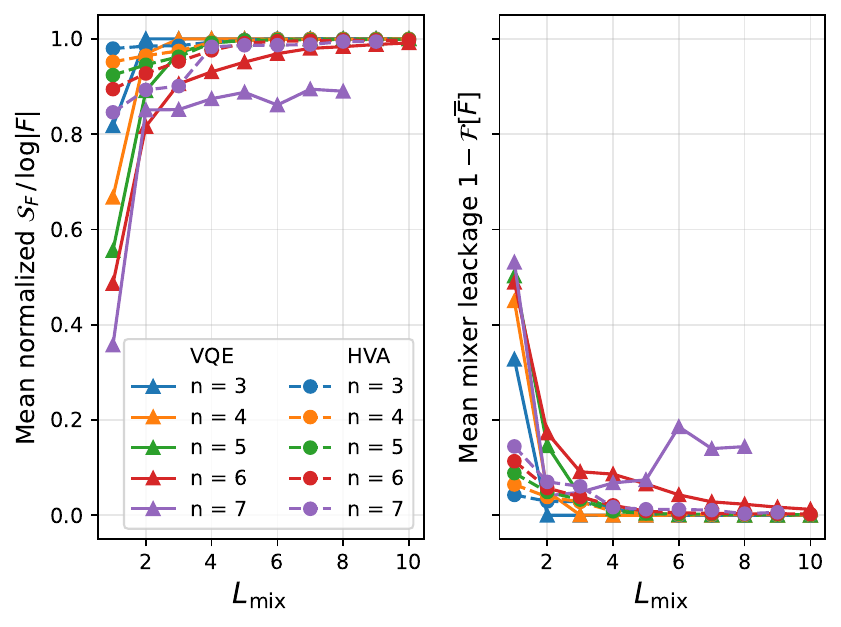}
    \caption{Average normalized entropy $\mathcal{S}_F/\log |F|$ (left panel) and leakage $1 - \mathcal{F}$ (right panel) of the initial state $|\overline{F}\rangle$ for the MWIS problem computed using VQE (full line with triangles) and HVA (dashed line with circles) as a function of the number of layers $L$ for different number of qubits $n$.}
    \label{fig:mis_learning}
\end{figure}

The MWIS problem consists of finding an independent set of vertices in a graph \(G=(V,E)\) that maximizes the total vertex weight~\cite{Diestel2025}. Given weights \(w_i\) for each vertex \(i\in V\), the problem can be formulated as
\begin{align}
&\min_{x \in \{0, 1\}^n}  \left(-\sum_{i\in V} w_i x_i\right),\\
&\text{subject to} \quad x_i+x_j\leq 1, \quad \forall (i,j)\in E \notag
\end{align}
where \(x_i=1\) if vertex \(i\) belongs to the independent set and \(x_i=0\) otherwise. The constraints ensure that no two adjacent vertices are simultaneously selected, while the objective maximizes the total weight of the selected vertices.
The instances are generated from random geometric graphs constructed using NetworkX~\cite{hagberg2008exploring}, with additional edges added to ensure graph connectivity. The vertex weights $w_i$ are independently sampled from a discrete uniform distribution over $\{1, \, 2,\ldots,6\}$. 

A penalty term to represent the constraint of MWIS is $x_i x_j$ for each edge $(i,j) \in E$. This avoids the need for slack variables, thus $Q=n$. Summing over all edges and mapping $x_i \to (\mathbb{1} - Z_i)/2$ gives the penalty Hamiltonian of the QUBO formulation,
\begin{align}
    \mathcal{H}_g = \frac{1}{4}\sum_{(i,j) \in E} (\mathbb{1} - Z_i)(\mathbb{1} - Z_j),
\end{align}
which vanishes on independent sets and is positive on all other bitstrings. Its ground space is therefore spanned by the feasible states, so the above $\mathcal{H}_g$ is a natural choice to construct the HVA with a single parameter per layer in the cost, i.e., $\mathcal{H}_C\left(\boldsymbol{\theta}_{C, k}\right) = \theta_{C, k}\mathcal{H}_g$ in Eq.~\eqref{eq:U_F_qaoa}.

\begin{figure}[t]
    \centering
    \includegraphics[trim={0 1.8cm 0 0}, width=\linewidth]{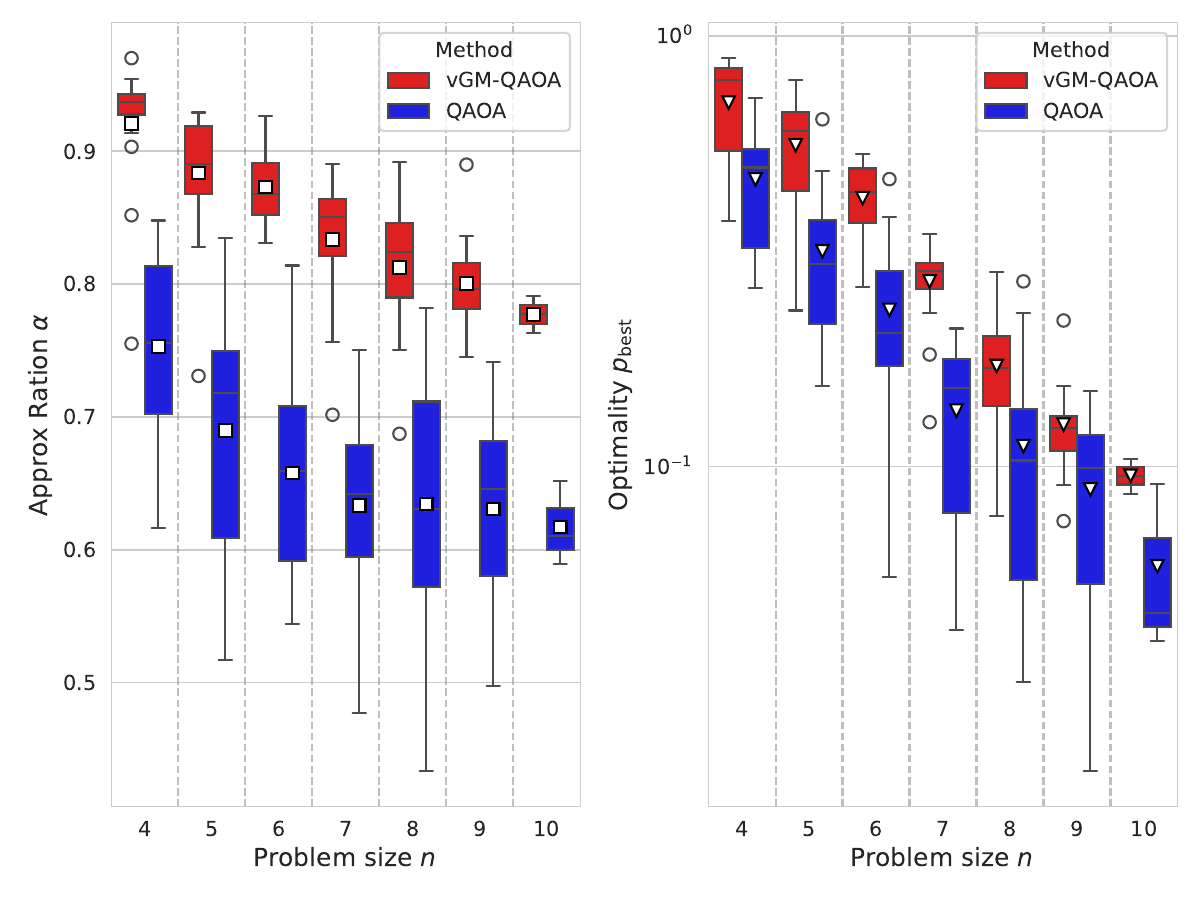}
    \caption{Approximation ratio $\alpha$ (left) and probability of the optimal state $p_{\rm best}$ (right) averaged over the MWIS instances as a function of the problem size $n$. vGM-QAOA is indicated in red, while QAOA is indicated in blue. The number of layers is $L=4$ and $L=6$ for vGM-QAOA and QAOA respectively. For vGM-QAOA, we use the QAOA-based $\overline{U}_F$ from Eq.~\eqref{eq:U_F_qaoa}.}
    \label{fig:mis_result}
\end{figure}

\subsubsection{Learning initial state}

In this section, we discuss the learning step of the state-preparation circuit $\overline{U}_F$ for the MWIS problem. Figure~\ref{fig:mis_learning} compares the VQE and HVA Ans\"atze as a function of the number of layers $L_{\rm mix}$, for different problem sizes $n$. For each configuration, we report the normalized entropy $\mathcal{S}_F/\log|F|$ and the leakage $1-\mathcal{F}[\overline{F}]$, averaged over $16$ instances. The plot shows convergence of both metrics as $L_{\rm mix}$ is increased. For a small number of qubits $n=3$ and $n=4$, VQE performs better than HVA. Above that threshold, HVA becomes more efficient at learning the state, in spite of its smaller number of parameters. This might be attributed to the fact that VQE suffers from barren plateaus and its training becomes more difficult as the number of qubits increases. Therefore, in the following we use the $\overline{U}^{\rm HVA}_F$ within vGM-QAOA for the MWIS problem.

\subsubsection{Performance of vGM-QAOA}

We now compare the full vGM-QAOA, built on the learned $\overline{U}_F$ from Eq.~\eqref{eq:U_F_qaoa}, with standard QUBO-based QAOA in terms of the approximation ratio $\alpha$ and the optimal-solution probability $p_{\rm best}$, for different problem sizes $n$. The results are collected in Fig.~\ref{fig:mis_result}. We recall that both algorithms use the same linear-ramp schedule and are optimized with the same loss $\mathcal{L}_F$, so that the comparison isolates the effect of the mixer.

Moreover, to ensure a fair comparison, we choose the number of layers $L_{\rm mix}$ of $\overline{U}_F$ such that the two approaches have comparable circuit depth once both are decomposed into single-qubit rotations and CX gates. Under this constraint, vGM-QAOA consistently exhibits improved performance relative to standard QAOA, yielding higher values of both $\alpha$ and $p_{\rm best}$. The probability of the best state still decreases exponentially as a function of $n$, indicating that only approximate high-quality solutions can be obtained at large scale.

\subsection{Multi-resource Knapsack}

\begin{figure}[t]
    \centering \includegraphics[trim={0 1.2cm 0 0}, width=\linewidth]{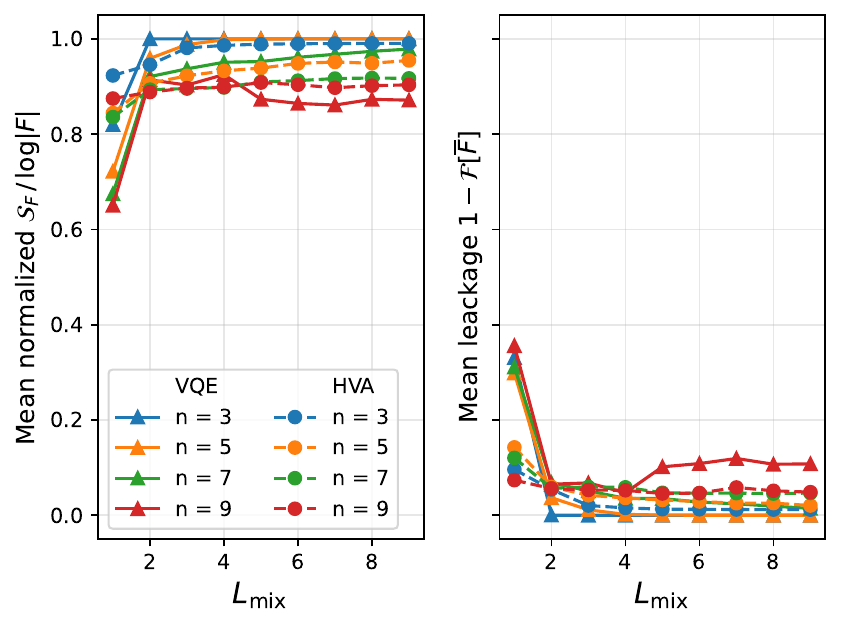}
    \caption{Average normalized entropy $\mathcal{S}_F/\log|F|$ (left) and leakage $1 - \mathcal{F}$ (right) of the initial state $|\overline{F}\rangle$ for the knapsack problem, obtained with $\overline{U}^{\rm VQE}_F$ and with $\overline{U}^{\rm HVA}_F$ using $\mathcal{H}_C$ of Eq.~\eqref{eq:mkp_up}, as a function of the number of layers $L_{\rm mix}$ for different numbers of qubits $n$. Averages are taken over 16 instances; optimizer details are given in the main text.}
    \label{fig:knapsack_learning}
\end{figure}

\begin{figure}[t]
    \centering
    \includegraphics[trim={0 1.2cm 0 0}, width=\linewidth]{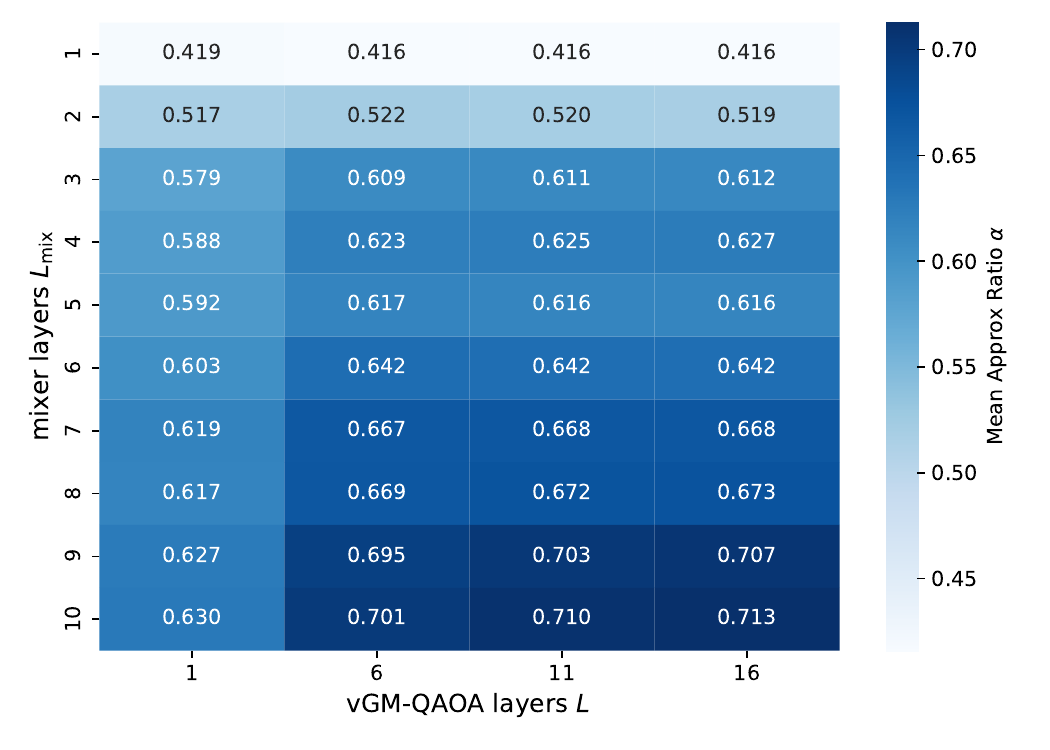}
    \caption{Heatmap of the average approximation ratio $\alpha$ over 16 instances of the Knapsack problem with $n=7$ as a function of the mixer layers $L_{\rm mix}$ and vGM-QAOA layers $L$. vGM-QAOA relies on $\overline{U}^{\rm VQE}_F$ for the state preparation and the mixer. Darker blue indicates higher $\alpha$ corresponding to better performance. }
    \label{fig:knapsack_heatmap}
\end{figure} 

\begin{figure}[t]
    \centering
    \includegraphics[trim={0 1.2cm 0 0}, width=\linewidth]{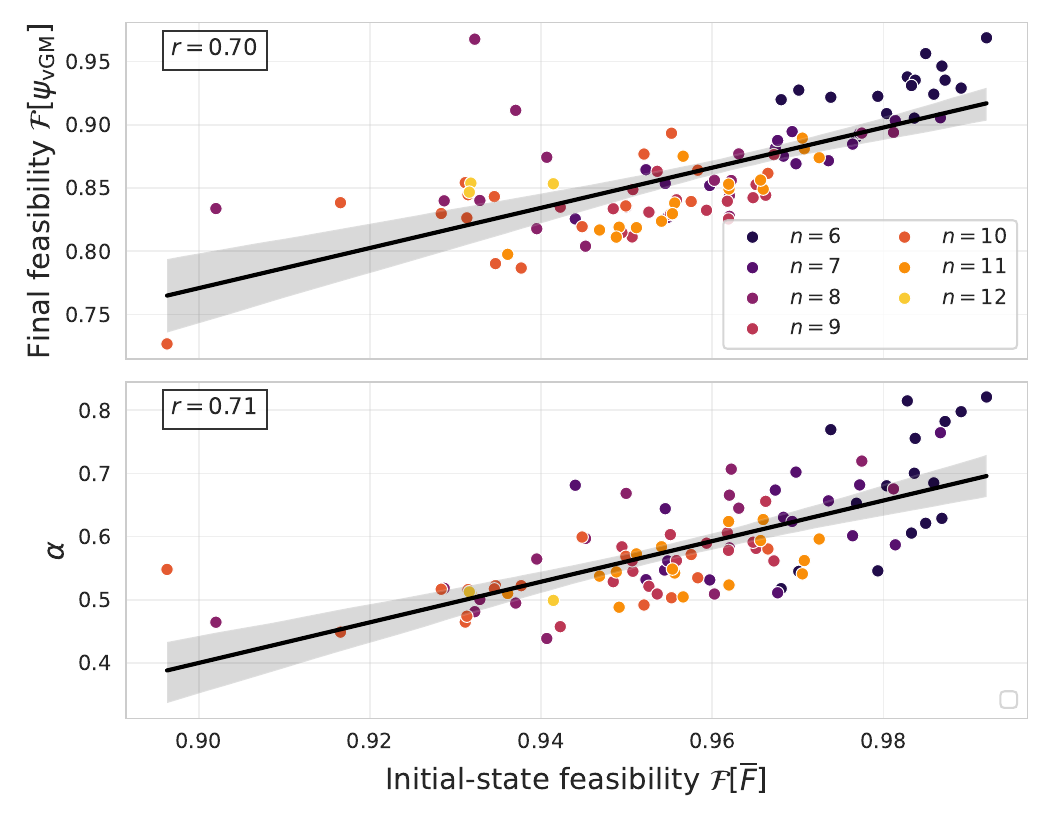}
    \caption{Final feasibility $\mathcal{F}[\psi_{\rm vGM}]$ (upper panel) and approximation ratio $\alpha$ (lower panel) of vGM-QAOA as a function of the initial-state feasibility $\mathcal{F}[\overline{F}]$, obtained with $\overline{U}^{\rm VQE}_F$. $L=6$ layers are used for vGM-QAOA, and $L_{\rm mix} = 3$ for the mixer circuit. Different values of the number of variables $n$ are represented by different colors. In each panel, the black line indicates the linear regression, and the shaded region represents its 95\% confidence interval. The boxes in the upper-left corners report the Pearson correlation coefficient $r$.}
    \label{fig:correl}
\end{figure}

The multi-resource Knapsack problem (MKP) consists of selecting a subset of items that maximizes the total value without exceeding a set of given capacity constraints. This problem has several industrial applications~\cite{Kellerer2004}. We consider $N$ items, each with value $v_i$, and a set $R$ of resources, each with maximum capacity $W_j \in \mathbb{Z}_{> 0}$. Selecting item $i$ consumes an amount $a_{ij} \in \mathbb{Z}_{\geq 0}$ of resource $j$. The goal is to select the subset of items with maximum total value that does not exceed any capacity:
\begin{align}
\label{eq:mkp_problem}
&\min_{x \in \{0, 1\}^n}  \left(- \sum_{i=1}^{n} v_i x_i\right) \\
&\text{subject to} \quad  \sum_{i=1}^{n} a_{ij} x_i \leq W_j, \forall j \in R \notag
\end{align}
Here, \(x_i=1\) if item \(i\) is selected and \(x_i=0\) otherwise. The capacity constraints ensure that, for each resource $j$, the total consumption of the selected items does not exceed the capacity $W_j$. 

In the standard QUBO formulation, each inequality constraint is first converted into an equality by introducing binary slack variables~\cite{glover2022qubo}. The total number
of necessary slack variables is
\begin{equation}
n_s
=
\sum_{j=1}^{|R|}
\left\lceil \log_2(W_j+1) \right\rceil.
\end{equation}
This expression assumes a conventional binary encoding; alternative encodings may require a different number of binary variables.

In this work, we consider MKP with $|R| = 1$ (single resource) and $|R| = 2$. Instances are randomly generated by drawing $v_i$ and $a_{ij}$ independently and uniformly from $\{1, \dots, 5\}$. Each capacity is set to half of the total consumption of the corresponding resource, $W_j = \lceil \frac{1}{2} \sum_{i=1}^{n} a_{ij} \rceil$, so that $W_j$, and with it $n_s$, grows with the number of items $n$. The resulting total number of qubits, $Q = n + n_s$, can severely limit the applicability of the QUBO approach, e.g. in classical emulation, especially for multiple resources.

The same overhead makes learning an effective mixer more challenging for MKP than for MWIS. In principle, the strategy adopted for MWIS could be extended by encoding the capacity constraints as QUBO penalty terms in $\mathcal{H}_C$, but the resulting state-preparation circuit would then act on all $Q$ qubits. We therefore focus on approaches that avoid slack variables altogether, so that $\overline{U}_F$ acts only on the $n$ decision variables. Specifically, we investigate two variational approaches: the VQE circuit of Eq.~\eqref{eq:U_F_vqe} and HVA circuit in the form of Eq.~\eqref{eq:U_F_qaoa} inspired by the unbalanced-penalty QAOA proposed in~\cite{Montañez-Barrera_2024}. For the latter circuit, we introduce the following expression for $\mathcal{H}_C$ in Eq.~\eqref{eq:U_F_qaoa}
\begin{align}
&\mathcal{H}_{C}\left(\boldsymbol{\theta}_{C,k}\right) = \sum_{j \in R}\left[\theta_{1,j;k} \mathcal{H}_{C,j}  + \theta_{2,j; k} \left(\mathcal{H}_{C, j}\right)^2\right].
    \label{eq:mkp_up}
\end{align}
where ${\mathcal{H}_{C,j} = \left[\sum_{i=1}^n a_{ij} \frac{(\mathbb{1} - Z_i)}{2} - W_j \mathbb{1}  \right]}$. Each layer therefore has $n_c = 2|R|$ constraint parameters, so that the HVA Ansatz contains $(1+2|R|)L_{\rm mix}$ parameters in total.

\subsubsection{Learning initial state}

For a single resource, $|R| = 1$, the MKP of Eq.~\eqref{eq:mkp_problem} reduces to the standard binary knapsack problem, on which we focus to analyze the learning of the initial state.
Figure~\ref{fig:knapsack_learning} compares the HVA and VQE circuits in terms of the normalized entropy $\mathcal{S}_F/\log|F|$ and the leakage $1-\mathcal{F}[\overline{F}]$, as a function of the number of layers $L_{\rm mix}$. Our numerical results show that both approaches achieve high feasibility rates, indicating that they can effectively address the constraints of the MKP without requiring additional qubits for slack-variable encoding. Their relative performance, however, depends on the problem size. VQE appears to perform better for instances with a smaller number of qubits, while QAOA exhibits more favorable scaling as the number of qubits increases. This trend may depend on the choice of the optimizer for the optimization of the parameters. However, it suggests that the two approaches may be advantageous in different problem-size regimes. In particular, QAOA appears promising for larger instances, where the trainability issue due to barren plateaus of the VQE becomes more severe.

Since $\overline{U}^{\rm VQE}_F$ outperforms $\overline{U}^{\rm HVA}_F$ for $2 \leq L_{\rm mix} < 5$ over the range of $n$ considered here, for simplicity in the following we use it as the state-preparation unitary.

\subsubsection{Dependence on depth and initial-state feasibility}

We study the dependence of the approximation ratio $\alpha$ of the full vGM-QAOA on both the number of layers $L_{\rm mix}$ of the state-preparation circuit $\overline{U}_F$, which determines the quality of the learned mixer, and the number of vGM-QAOA layers $L$. The results are shown in Fig.~\ref{fig:knapsack_heatmap} for $n = 7$ variables.  The heatmap shows that, for any fixed $L$, $\alpha$ consistently improves as $L_{\rm mix}$ is increased. Similarly, at fixed $L_{\rm mix}$, $\alpha$ improves with increasing $L$, provided that the state-preparation circuit is deep enough, in this case for $L_{\rm mix} > 2$. Below this threshold, no clear trend can be established. This suggests that additional vGM-QAOA layers are beneficial only once $\overline{U}_F$ approximates the feasible superposition sufficiently well, whereas a poorly learned mixer cannot be compensated by a deeper outer circuit. Therefore, in the following, we set $L_{\rm mix} = 3$, ensuring that the vGM-QAOA and standard QAOA circuits have comparable depths while providing a sufficiently expressive state-preparation unitary.

Figure~\ref{fig:correl} complements this analysis by relating the quality of the learned initial state to the final performance, for instances of various sizes. Both the final feasibility $\mathcal{F}[\psi_{\rm vGM}]$ and the approximation ratio $\alpha$ clearly correlate with the initial-state feasibility $\mathcal{F}[\overline{F}]$, with an estimated Pearson coefficient around $r = 0.7$ indicating a strong positive linear correlation. In nearly all instances, the final feasibility is lower than that of the initial state, indicating that the residual leakage of $|\overline{F}\rangle$ is amplified by the repeated application of the approximate mixer. 
Moreover, we observe a decrease in performance at larger instances, which exhibit lower values of $\mathcal{F}[\overline{F}]$ and $\alpha$. This could originate, at least in part, from the difficulty of learning an accurate $\overline{U}_F$ for the initial state. Together with the heatmap, these results identify the accuracy of the learned state preparation as a key factor determining the performance of vGM-QAOA.

\begin{figure}
    \centering
    \includegraphics[trim={0 1.8cm 0 0}, width=\linewidth]{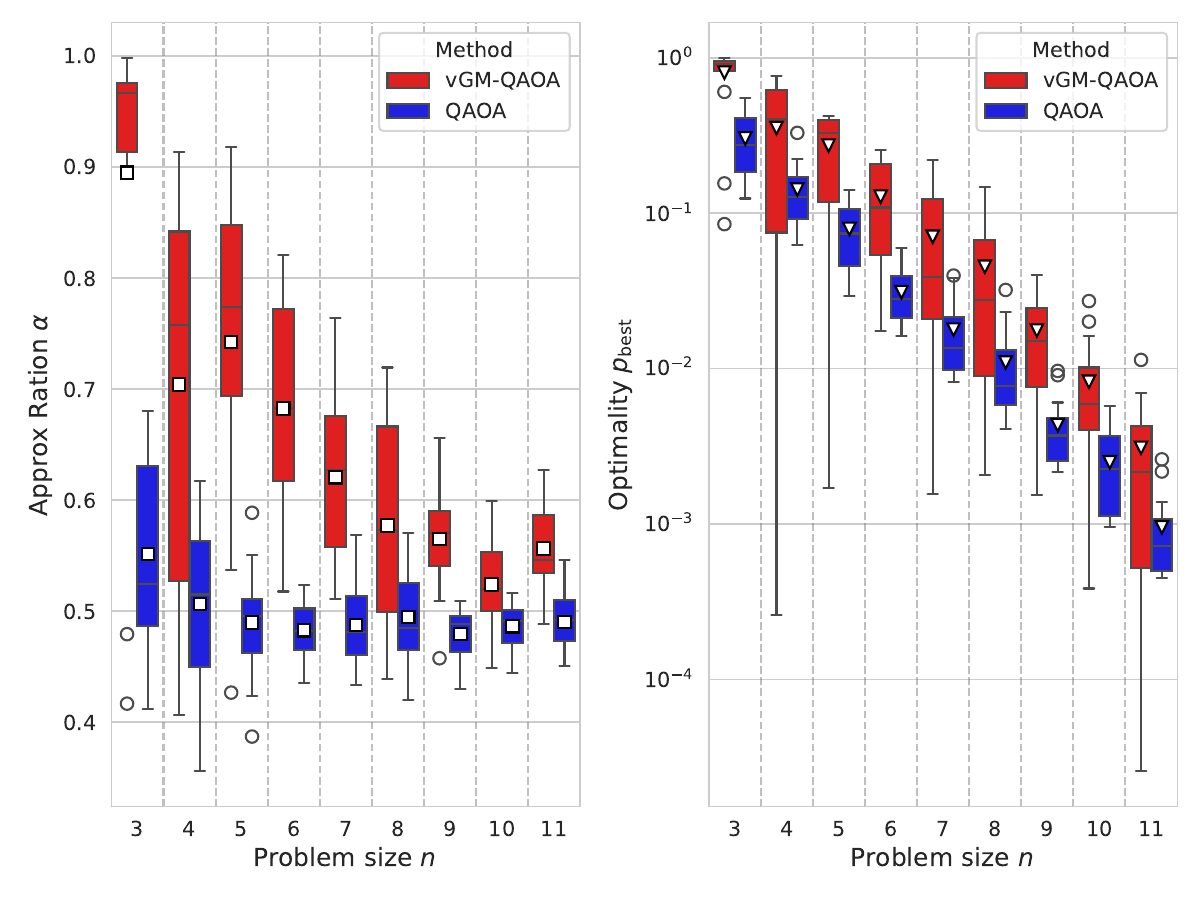}
    \caption{Approximation ratio $\alpha$ (left) and probability of the optimal state $p_{\rm best}$ (right) averaged over 16 random MKP instances with $R=1$ as a function of the problem size $n$. vGM-QAOA is indicated in red, while QAOA is indicated in blue. The number of layers is $L=6$ for both vGM-QAOA and QAOA. The layers of the VQE mixer are $L_{\rm mix} = 3$.}
    \label{fig:knapsack_result}
\end{figure}

\subsubsection{Performance of vGM-QAOA in the single- and multi-resource case}

Finally, we compare vGM-QAOA and standard QAOA in terms of $\alpha$ and $p_{\rm best}$ as a function of the problem size $n$, starting from the single-resource case, i.e., the standard knapsack problem. The results are shown in Fig.~\ref{fig:knapsack_result}, where the boxplots summarize both the average performance and its variability across instances. To ensure a fair comparison, as done for MWIS, we choose the number of layers $L_{\rm mix}$ such that vGM-QAOA has a circuit depth similar to that of QAOA, once both are decomposed into single-qubit rotations and CX gates.

Within this condition, we observe that vGM-QAOA generally outperforms standard QAOA, although the gap narrows as $n$ increases, and the two methods reach comparable values of $\alpha$ for the largest instances. This behavior is consistent with the correlation analysis of Fig.~\ref{fig:correl}: as $n$ grows, the learned $\overline{U}_F$ becomes harder to train and thus less accurate, which consequently reduces the benefit of the learned mixer. Nevertheless, vGM-QAOA requires fewer qubits, with $Q=n$, potentially enabling the solution of larger MKP instances than standard QAOA. This advantage becomes particularly relevant when the number of slack variables $n_s$ causes the qubit requirements of the QUBO formulation to exceed the limits of statevector emulation or current quantum hardware.

\begin{figure}[t]
    \centering
    \includegraphics[trim={0 1.8cm 0 0}, width=\linewidth]{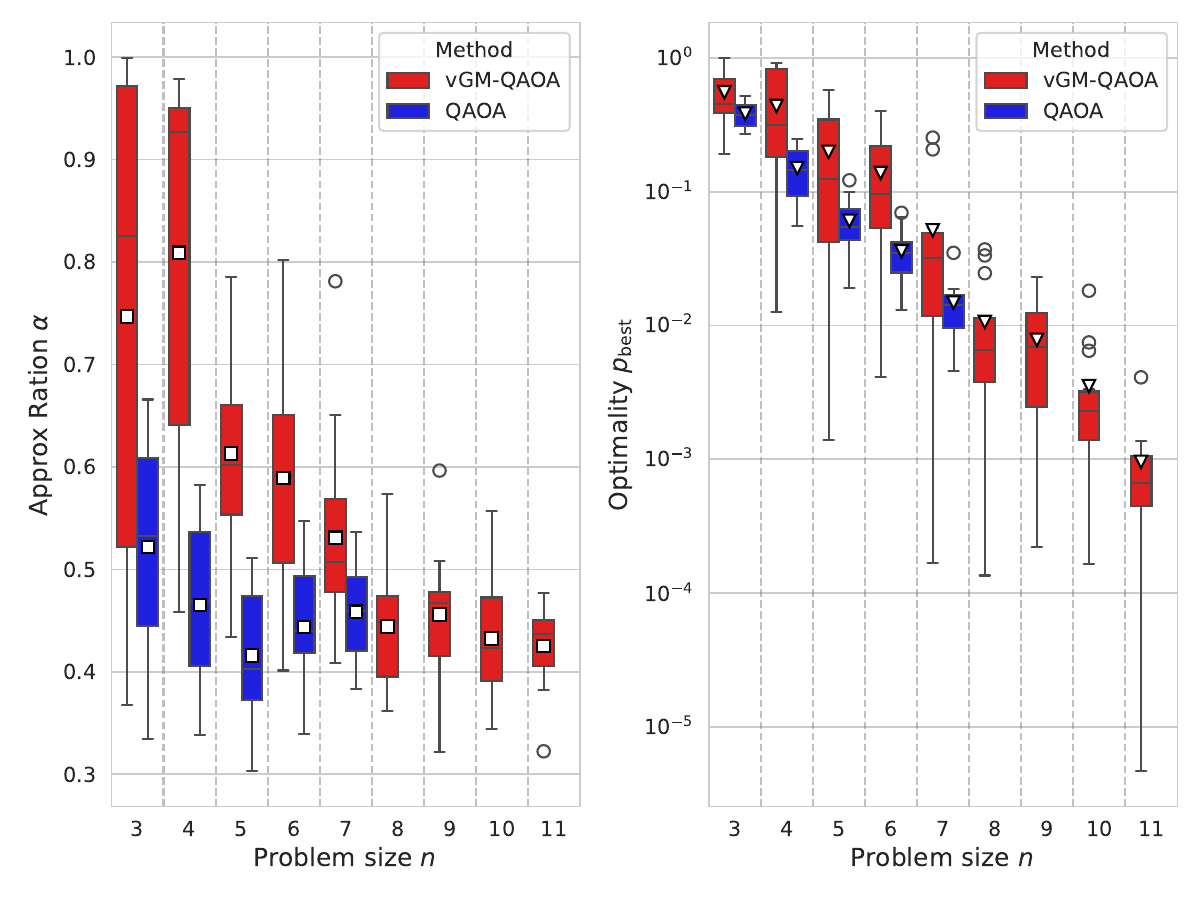}
    \caption{Approximation ratio $\alpha$ (left) and probability of measuring the optimal state $p_{\rm best}$ (right), averaged over 16 randomly generated MKP instances with two resources $(|R|=2)$, as functions of the problem size $n$. vGM-QAOA and standard QAOA are shown in red and blue, respectively. Both algorithms use $L=6$ layers, while the VQE mixer uses $L_{\rm mix}=3$ layers. For $n>7$, the number of qubits required by standard QAOA exceeded the practical limit for collecting statistics while performing the optimization iterations with the statevector emulator.}
    \label{fig:mkp_result}
\end{figure}

For completeness, we also compare vGM-QAOA and standard QAOA on MKP instances with $|R| = 2$ (Fig.~\ref{fig:mkp_result}). Here, the slack variables substantially increase the qubit count of the QUBO formulation, which already requires more than $\langle Q \rangle = 18$ qubits on average for $n = 7$. Since collecting statistics on top of the optimization then becomes computationally demanding with the statevector emulator, standard QAOA is limited to $n \leq 7$. vGM-QAOA avoids this overhead, since it acts on $Q = n$ qubits only, but its approximation ratio $\alpha$ remains relatively low for larger instances. This indicates that learning an accurate $\overline{U}_F$ remains the main challenge, in line with the single-resource results.

\section{Conclusions and Outlook}

In this work, we proposed vGM-QAOA, a variational extension of the Grover-mixer QAOA (GM-QAOA) for constrained combinatorial optimization. Instead of requiring an exact circuit for the uniform superposition of feasible states, which is generally unavailable, we train a parameterized quantum circuit to approximately prepare it, using an entropy-based cost function that favors a uniform distribution over the feasible subspace. The trained circuit then provides both the initial state and the mixer of GM-QAOA. We explored two circuit architectures for this task: a hardware-efficient VQE-type circuit and a Hamiltonian variational ansatz built from the problem constraints.

We evaluated the proposed framework on two benchmark problems with inequality constraints: the Maximum Weight Independent Set (MWIS) problem and the Multidimensional Knapsack Problem (MKP). Our results demonstrate the potential of the proposed approach to outperform standard QAOA, at least on small problem instances. Specifically, for MWIS, the mixer based on the learned $\overline{U}^{\rm HVA}_F$ consistently outperforms standard QAOA with the conventional $X$-mixer at comparable circuit depth, across all problem sizes considered. For MKP, the VQE-based mixer allows us to avoid the use of slack variables, but the advantage over QAOA becomes less significant as the problem size increases. This behavior appears to originate mainly from the increasing difficulty of training the state-preparation unitary $\overline{U}_F$ as the problem size grows.

The performance gains of vGM-QAOA must be weighed against the additional resources required to learn and implement $\overline{U}_F$. In particular, increasing the depth of standard QAOA may partially or entirely close the performance gap, potentially at a lower computational cost. The practical advantage of vGM-QAOA therefore depends on the trade-off between the quality of the learned mixer, the resources required for its training, and the performance achievable by standard QAOA.

Future work will focus on improving the training of the state-preparation circuit, which we identified as the main bottleneck of the method, in particular to reduce the leakage into infeasible states that the mixer amplifies. Assessing the effects of finite-shot sampling and hardware noise is also a necessary step towards the practical applicability of the method.

\section*{Acknowledgments}
The authors would like to thank the whole Quantum Computing Solutions group at Leonardo S.p.A. for useful discussions.

\section*{Disclaimer}
The authors declare no competing interest.

\bibliographystyle{unsrturl}
\bibliography{bibliog}

@article{petropoulos2024or,
author = 
{Fotios Petropoulos and  Gilbert Laporte and  Emel Aktas and  Sibel A. Alumur and  Claudia Archetti and  Hayriye Ayhan and  Maria Battarra and  Julia A. Bennell and  Jean-Marie Bourjolly and  John E. Boylan and  Michèle Breton and  David Canca and  Laurent Charlin and  Bo Chen and  Cihan Tugrul Cicek and  Louis Anthony Cox Jr and  Christine S.M. Currie and  Erik Demeulemeester and  Li Ding and  Stephen M. Disney and  Matthias Ehrgott and  Martin J. Eppler and  Güneş Erdoğan and  Bernard Fortz and  L. Alberto Franco and  Jens Frische and  Salvatore Greco and  Amanda J. Gregory and  Raimo P. Hämäläinen and  Willy Herroelen and  Mike Hewitt and  Jan Holmström and  John N. Hooker and  Tuğçe Işık and  Jill Johnes and  Bahar Y. Kara and  Özlem Karsu and  Katherine Kent and  Charlotte Köhler and  Martin Kunc and  Yong-Hong Kuo and  Adam N. Letchford and  Janny Leung and  Dong Li and  Haitao Li and  Judit Lienert and  Ivana Ljubić and  Andrea Lodi and  Sebastián Lozano and  Virginie Lurkin and  Silvano Martello and  Ian G. McHale and  Gerald Midgley and  John D.W. Morecroft and  Akshay Mutha and  Ceyda Oğuz and  Sanja Petrovic and  Ulrich Pferschy and  Harilaos N. Psaraftis and  Sam Rose and  Lauri Saarinen and  Said Salhi and  Jing-Sheng Song and  Dimitrios Sotiros and  Kathryn E. Stecke and  Arne K. Strauss and  İstenç Tarhan and  Clemens Thielen and  Paolo Toth and  Tom Van Woensel and  Greet Vanden Berghe and  Christos Vasilakis and  Vikrant Vaze and  Daniele Vigo and  Kai Virtanen and  Xun Wang and  Rafał Weron and  Leroy White and  Mike Yearworth and  E. Alper Yıldırım and  Georges Zaccour and Xuying Zhao},
title = {Operational Research: methods and applications},
journal = {Journal of the Operational Research Society},
volume = {75},
number = {3},
pages = {423--617},
year = {2024},
publisher = {Taylor \& Francis},
doi = {10.1080/01605682.2023.2253852},
}

@inbook{nemhauser1988integer,
author = {Nemhauser, George L. and Wolsey, Laurence A.},
title = {Integer and combinatorial optimization},
year = {1988},
isbn = {047182819X},
publisher = {Wiley-Interscience},
address = {USA},
doi = {https://doi.org/10.1002/9781118627372},
}

@article{land1960bb,
  author    = {Ailsa H. Land and Alison G. Doig},
  title     = {An automatic method of solving discrete programming problems},
  journal   = {Econometrica},
  volume    = {28},
  number    = {3},
  pages     = {497--520},
  year      = {1960},
  publisher = {The Econometric Society},
  doi       = {10.2307/1910129}
}

@misc{farhi2000quantum,
      title={Quantum Computation by Adiabatic Evolution}, 
      author={Edward Farhi and Jeffrey Goldstone and Sam Gutmann and Michael Sipser},
      year={2000},
      eprint={quant-ph/0001106},
      archivePrefix={arXiv},
      primaryClass={quant-ph}
}

@Article{lucas2014ising,
author={Lucas, Andrew}, 
title={Ising formulations of many NP problems}, 
journal={Frontiers in Physics},     
volume={2},
year={2014},
url={https://www.frontiersin.org/journals/physics/articles/10.3389/fphy.2014.00005},
doi={10.3389/fphy.2014.00005}
}

@Article{glover2022qubo,
author={Glover, Fred
and Kochenberger, Gary
and Hennig, Rick
and Du, Yu},
title={Quantum bridge analytics I: a tutorial on formulating and using QUBO models},
journal={Annals of Operations Research},
year={2022},
month={Jul},
day={01},
volume={314},
number={1},
pages={141-183},
issn={1572-9338},
doi={10.1007/s10479-022-04634-2},
url={https://doi.org/10.1007/s10479-022-04634-2}
}

@Article{hadfield2019qaoa,
AUTHOR = {Hadfield, Stuart and Wang, Zhihui and O’Gorman, Bryan and Rieffel, Eleanor G. and Venturelli, Davide and Biswas, Rupak},
TITLE = {From the Quantum Approximate Optimization Algorithm to a Quantum Alternating Operator Ansatz},
JOURNAL = {Algorithms},
VOLUME = {12},
YEAR = {2019},
NUMBER = {2},
ARTICLE-NUMBER = {34},
URL = {https://www.mdpi.com/1999-4893/12/2/34},
ISSN = {1999-4893},
DOI = {10.3390/a12020034}
}

@article{PhysRevA.101.012320,
  title = {$XY$ mixers: Analytical and numerical results for the quantum alternating operator ansatz},
  author = {Wang, Zhihui and Rubin, Nicholas C. and Dominy, Jason M. and Rieffel, Eleanor G.},
  journal = {Phys. Rev. A},
  volume = {101},
  issue = {1},
  pages = {012320},
  numpages = {16},
  year = {2020},
  month = {Jan},
  publisher = {American Physical Society},
  doi = {10.1103/PhysRevA.101.012320},
  url = {https://link.aps.org/doi/10.1103/PhysRevA.101.012320}
}

@INPROCEEDINGS{baertschi2020grover,
  author={Bärtschi, Andreas and Eidenbenz, Stephan},
  booktitle={2020 IEEE International Conference on Quantum Computing and Engineering (QCE)}, 
  title={Grover Mixers for QAOA: Shifting Complexity from Mixer Design to State Preparation}, 
  year={2020},
  volume={},
  number={},
  pages={72-82},
  doi={10.1109/QCE49297.2020.00020}}

@article{BOUSSAID201382,
title = {A survey on optimization metaheuristics},
journal = {Information Sciences},
volume = {237},
pages = {82-117},
year = {2013},
note = {Prediction, Control and Diagnosis using Advanced Neural Computations},
issn = {0020-0255},
doi = {https://doi.org/10.1016/j.ins.2013.02.041},
url = {https://www.sciencedirect.com/science/article/pii/S0020025513001588},
author = {Ilhem Boussaïd and Julien Lepagnot and Patrick Siarry}
}

@Article{montanez2025linear,
author={Monta{\~{n}}ez-Barrera, J. A.
and Michielsen, Kristel},
title={Toward a linear-ramp QAOA protocol: evidence of a scaling advantage in solving some combinatorial optimization problems},
journal={npj Quantum Information},
year={2025},
month={Aug},
day={04},
volume={11},
number={1},
pages={131},
issn={2056-6387},
doi={10.1038/s41534-025-01082-1},
url={https://doi.org/10.1038/s41534-025-01082-1}
}

@article{blekos2024qaoa,
   title={A review on Quantum Approximate Optimization Algorithm and its variants},
   volume={1068},
   ISSN={0370-1573},
   url={http://dx.doi.org/10.1016/j.physrep.2024.03.002},
   DOI={10.1016/j.physrep.2024.03.002},
   journal={Physics Reports},
   publisher={Elsevier BV},
   author={Blekos, Kostas and Brand, Dean and Ceschini, Andrea and Chou, Chiao-Hui and Li, Rui-Hao and Pandya, Komal and Summer, Alessandro},
   year={2024},
   month=jun, pages={1–66} }

@misc{farhi2014quantumapproximateoptimizationalgorithm,
      title={A Quantum Approximate Optimization Algorithm}, 
      author={Edward Farhi and Jeffrey Goldstone and Sam Gutmann},
      year={2014},
      eprint={1411.4028},
      archivePrefix={arXiv},
      primaryClass={quant-ph},
      url={https://arxiv.org/abs/1411.4028}, 
}

@misc{javadiabhari2024quantumcomputingqiskit,
      title={Quantum computing with Qiskit}, 
      author={Ali Javadi-Abhari and Matthew Treinish and Kevin Krsulich and Christopher J. Wood and Jake Lishman and Julien Gacon and Simon Martiel and Paul D. Nation and Lev S. Bishop and Andrew W. Cross and Blake R. Johnson and Jay M. Gambetta},
      year={2024},
      eprint={2405.08810},
      archivePrefix={arXiv},
      primaryClass={quant-ph},
      url={https://arxiv.org/abs/2405.08810}, 
}

@Article{abbas2024review,
author={Abbas, Amira
and Ambainis, Andris
and Augustino, Brandon
and B{\"a}rtschi, Andreas
and Buhrman, Harry
and Coffrin, Carleton
and Cortiana, Giorgio
and Dunjko, Vedran
and Egger, Daniel J.
and Elmegreen, Bruce G.
and Franco, Nicola
and Fratini, Filippo
and Fuller, Bryce
and Gacon, Julien
and Gonciulea, Constantin
and Gribling, Sander
and Gupta, Swati
and Hadfield, Stuart
and Heese, Raoul
and Kircher, Gerhard
and Kleinert, Thomas
and Koch, Thorsten
and Korpas, Georgios
and Lenk, Steve
and Marecek, Jakub
and Markov, Vanio
and Mazzola, Guglielmo
and Mensa, Stefano
and Mohseni, Naeimeh
and Nannicini, Giacomo
and O'Meara, Corey
and Tapia, Elena Pe{\~{n}}a
and Pokutta, Sebastian
and Proissl, Manuel
and Rebentrost, Patrick
and Sahin, Emre
and Symons, Benjamin C. B.
and Tornow, Sabine
and Valls, V{\'i}ctor
and Woerner, Stefan
and Wolf-Bauwens, Mira L.
and Yard, Jon
and Yarkoni, Sheir
and Zechiel, Dirk
and Zhuk, Sergiy
and Zoufal, Christa},
title={Challenges and opportunities in quantum optimization},
journal={Nature Reviews Physics},
year={2024},
month={Dec},
day={01},
volume={6},
number={12},
pages={718-735},
issn={2522-5820},
doi={10.1038/s42254-024-00770-9},
url={https://doi.org/10.1038/s42254-024-00770-9}
}

@inproceedings{hagberg2008exploring,
  author    = {Aric A. Hagberg and Daniel A. Schult and Pieter J. Swart},
  title     = {Exploring Network Structure, Dynamics, and Function using {NetworkX}},
  booktitle = {Proceedings of the 7th Python in Science Conference (SciPy 2008)},
  editor    = {Ga{\"e}l Varoquaux and Travis Vaught and Jarrod Millman},
  pages     = {11--15},
  address   = {Pasadena, CA, USA},
  month     = aug,
  year      = {2008}
}

@article{mcclean2018barren,
  author    = {McClean, Jarrod R. and Boixo, Sergio and Smelyanskiy, Vadim N. and Babbush, Ryan and Neven, Hartmut},
  title     = {Barren plateaus in quantum neural network training landscapes},
  journal   = {Nature Communications},
  volume    = {9},
  number    = {1},
  pages     = {4812},
  year      = {2018},
  publisher = {Nature Publishing Group},
  doi       = {10.1038/s41467-018-07090-4}
}

@Article{Ragone2024,
author={Ragone, Michael
and Bakalov, Bojko N.
and Sauvage, Fr{\'e}d{\'e}ric
and Kemper, Alexander F.
and Ortiz Marrero, Carlos
and Larocca, Mart{\'i}n
and Cerezo, M.},
title={A Lie algebraic theory of barren plateaus for deep parameterized quantum circuits},
journal={Nature Communications},
year={2024},
month={Aug},
day={22},
volume={15},
number={1},
pages={7172},
issn={2041-1723},
doi={10.1038/s41467-024-49909-3},
url={https://doi.org/10.1038/s41467-024-49909-3}
}

@article{Montañez-Barrera_2024,
doi = {10.1088/2058-9565/ad35e4},
url = {https://doi.org/10.1088/2058-9565/ad35e4},
year = {2024},
month = {apr},
publisher = {IOP Publishing},
volume = {9},
number = {2},
pages = {025022},
author = {Montañez-Barrera, J A and Willsch, Dennis and Maldonado-Romo, A and Michielsen, Kristel},
title = {Unbalanced penalization: a new approach to encode inequality constraints of combinatorial problems for quantum optimization algorithms},
journal = {Quantum Science and Technology}
}

@book{papadimitriou1982combinatorial,
author = {Papadimitriou, Christos H. and Steiglitz, Kenneth},
title = {Combinatorial optimization: algorithms and complexity},
year = {1982},
isbn = {0131524623},
publisher = {Prentice-Hall, Inc.},
address = {USA}
}

@article{peruzzo2014variational,
  author    = {Peruzzo, Alberto and McClean, Jarrod and Shadbolt, Peter and Yung, Man-Hong and Zhou, Xiao-Qi and Love, Peter J. and Aspuru-Guzik, Al{\'a}n and O'Brien, Jeremy L.},
  title     = {A variational eigenvalue solver on a photonic quantum processor},
  journal   = {Nature Communications},
  volume    = {5},
  number    = {1},
  pages     = {4213},
  year      = {2014},
  publisher = {Nature Publishing Group},
  doi       = {10.1038/ncomms5213}
}

@article{kandala2017hardware,
  author    = {Kandala, Abhinav and Mezzacapo, Antonio and Temme, Kristan and Takita, Maika and Brink, Markus and Chow, Jerry M. and Gambetta, Jay M.},
  title     = {Hardware-efficient variational quantum eigensolver for small molecules and quantum magnets},
  journal   = {Nature},
  volume    = {549},
  number    = {7671},
  pages     = {242--246},
  year      = {2017},
  publisher = {Nature Publishing Group},
  doi       = {10.1038/nature23879}
}

@article{cerezo2021variational,
  author    = {Cerezo, M. and Arrasmith, Andrew and Babbush, Ryan and Benjamin, Simon C. and Endo, Suguru and Fujii, Keisuke and McClean, Jarrod R. and Mitarai, Kosuke and Yuan, Xiao and Cincio, Lukasz and Coles, Patrick J.},
  title     = {Variational quantum algorithms},
  journal   = {Nature Reviews Physics},
  volume    = {3},
  number    = {9},
  pages     = {625--644},
  year      = {2021},
  publisher = {Nature Publishing Group},
  doi       = {10.1038/s42254-021-00348-9}
}

@misc{gabbassov2024lagrangian,
      title={Lagrangian Duality in Quantum Optimization: Overcoming QUBO Limitations for Constrained Problems}, 
      author={Einar Gabbassov and Gili Rosenberg and Artur Scherer},
      year={2024},
      eprint={2310.04542},
      archivePrefix={arXiv},
      primaryClass={math.OC},
      url={https://arxiv.org/abs/2310.04542}, 
}

@ARTICLE{shirai2025compressed,
  author={Shirai, Tatsuhiko and Togawa, Nozomu},
  journal={IEEE Transactions on Quantum Engineering}, 
  title={Compressed Space Quantum Approximate Optimization Algorithm for Constrained Combinatorial Optimization}, 
  year={2025},
  volume={6},
  number={},
  pages={1-14},
  doi={10.1109/TQE.2025.3602404}}

@Article{         harris2020array,
 title         = {Array programming with {NumPy}},
 author        = {Charles R. Harris and K. Jarrod Millman and St{\'{e}}fan J.
                 van der Walt and Ralf Gommers and Pauli Virtanen and David
                 Cournapeau and Eric Wieser and Julian Taylor and Sebastian
                 Berg and Nathaniel J. Smith and Robert Kern and Matti Picus
                 and Stephan Hoyer and Marten H. van Kerkwijk and Matthew
                 Brett and Allan Haldane and Jaime Fern{\'{a}}ndez del
                 R{\'{i}}o and Mark Wiebe and Pearu Peterson and Pierre
                 G{\'{e}}rard-Marchant and Kevin Sheppard and Tyler Reddy and
                 Warren Weckesser and Hameer Abbasi and Christoph Gohlke and
                 Travis E. Oliphant},
 year          = {2020},
 month         = sep,
 journal       = {Nature},
 volume        = {585},
 number        = {7825},
 pages         = {357--362},
 doi           = {10.1038/s41586-020-2649-2},
 publisher     = {Springer Science and Business Media {LLC}},
 url           = {https://doi.org/10.1038/s41586-020-2649-2}
}

@article{SZU1987157,
title = {Fast simulated annealing},
journal = {Physics Letters A},
volume = {122},
number = {3},
pages = {157-162},
year = {1987},
issn = {0375-9601},
doi = {https://doi.org/10.1016/0375-9601(87)90796-1},
url = {https://www.sciencedirect.com/science/article/pii/0375960187907961},
author = {Harold Szu and Ralph Hartley}
}

@ARTICLE{2020SciPy-NMeth,
  author  = {Virtanen, Pauli and Gommers, Ralf and Oliphant, Travis E. and
            Haberland, Matt and Reddy, Tyler and Cournapeau, David and
            Burovski, Evgeni and Peterson, Pearu and Weckesser, Warren and
            Bright, Jonathan and {van der Walt}, St{\'e}fan J. and
            Brett, Matthew and Wilson, Joshua and Millman, K. Jarrod and
            Mayorov, Nikolay and Nelson, Andrew R. J. and Jones, Eric and
            Kern, Robert and Larson, Eric and Carey, C J and
            Polat, {\.I}lhan and Feng, Yu and Moore, Eric W. and
            {VanderPlas}, Jake and Laxalde, Denis and Perktold, Josef and
            Cimrman, Robert and Henriksen, Ian and Quintero, E. A. and
            Harris, Charles R. and Archibald, Anne M. and
            Ribeiro, Ant{\^o}nio H. and Pedregosa, Fabian and
            {van Mulbregt}, Paul and {SciPy 1.0 Contributors}},
  title   = {{{SciPy} 1.0: Fundamental Algorithms for Scientific
            Computing in Python}},
  journal = {Nature Methods},
  year    = {2020},
  volume  = {17},
  pages   = {261--272},
  adsurl  = {https://rdcu.be/b08Wh},
  doi     = {10.1038/s41592-019-0686-2},
}

@misc{gabbassov2026principles,
      title={Principles of Quantum Optimization for Constrained Problems}, 
      author={Einar Gabbassov and Gurpahul Singh and Achim Kempf},
      year={2026},
      eprint={2607.14227},
      archivePrefix={arXiv},
      primaryClass={quant-ph},
      url={https://arxiv.org/abs/2607.14227}, 
}

@Article{a15060202,
AUTHOR = {Fuchs, Franz Georg and Lye, Kjetil Olsen and Møll Nilsen, Halvor and Stasik, Alexander Johannes and Sartor, Giorgio},
TITLE = {Constraint Preserving Mixers for the Quantum Approximate Optimization Algorithm},
JOURNAL = {Algorithms},
VOLUME = {15},
YEAR = {2022},
NUMBER = {6},
ARTICLE-NUMBER = {202},
URL = {https://www.mdpi.com/1999-4893/15/6/202},
ISSN = {1999-4893},
DOI = {10.3390/a15060202}
}

@misc{tsvelikhovskiy2026expressive,
      title={The Expressive Power of Constrained QAOA: What You Might Have MISsed}, 
      author={Boris Tsvelikhovskiy and Bao Bach and Ilya Safro},
      year={2026},
      eprint={2609.18209},
      archivePrefix={arXiv},
      primaryClass={quant-ph},
      url={https://arxiv.org/abs/2609.18209}, 
}

@misc{vandelli2025constraint,
      title={Constraint-preserving quantum algorithm for the multi-frequency antenna placement problem}, 
      author={Matteo Vandelli and Francesco Ferrari and Daniele Dragoni},
      year={2025},
      eprint={2511.15566},
      archivePrefix={arXiv},
      primaryClass={quant-ph},
      url={https://arxiv.org/abs/2511.15566}, 
}

@misc{awasthi2026constraint,
      title={Constraint Preserving XY-Mixers under Trotterized Adiabatic Evolution}, 
      author={Abhishek Awasthi and Maximilian Hess and Salome Lomadze and Francesco Bär and Christian Biefel},
      year={2026},
      eprint={2605.02465},
      archivePrefix={arXiv},
      primaryClass={quant-ph},
      url={https://arxiv.org/abs/2605.02465}, 
}

@misc{note_phases,
  note = {Suppose $|\overline{F}\rangle$ differs from $|F\rangle$ only by bitstring-dependent phases, namely it can be written as $|\overline{F}\rangle = D|F\rangle$, with $D = \sum_x e^{i\varphi_x}|x\rangle\langle x|$ a diagonal unitary in the computational basis. The approximate Grover mixer becomes $D\, U_M D^\dagger$. Since $D$ commutes with the diagonal cost unitary $e^{-i\gamma H_f}$, the final GM-QAOA state is $D|\psi_{\rm GM}\rangle$, which has the same measurement probabilities as $|\psi_{\rm GM}\rangle$. Therefore the relative phases do not affect the performance of GM-QAOA.}
}

@misc{ferrari2026feasibility,
      title={Feasibility-driven QAOA with penalty scheduling}, 
      author={Francesco Ferrari and Matteo Vandelli and Daniele Dragoni},
      year={2026},
      eprint={2606.25117},
      archivePrefix={arXiv},
      primaryClass={quant-ph},
      url={https://arxiv.org/abs/2606.25117}, 
}

@Inbook{Diestel2025,
author="Diestel, Reinhard",
title="The Basics",
bookTitle="Graph Theory",
year="2025",
publisher="Springer Berlin Heidelberg",
address="Berlin, Heidelberg",
pages="1--35",
isbn="978-3-662-70107-2",
doi="10.1007/978-3-662-70107-2_1",
url="https://doi.org/10.1007/978-3-662-70107-2_1"
}

@misc{zhou2026learned,
      title={Learned-projector QAOA for hierarchical optimization}, 
      author={Kangyun Zhou and Dong An and Jin-Peng Liu},
      year={2026},
      eprint={2609.28888},
      archivePrefix={arXiv},
      primaryClass={quant-ph},
      url={https://arxiv.org/abs/2609.28888}, 
}

@Inbook{Kellerer2004,
author="Kellerer, Hans
and Pferschy, Ulrich
and Pisinger, David",
title="Some Selected Applications",
bookTitle="Knapsack Problems",
year="2004",
publisher="Springer Berlin Heidelberg",
address="Berlin, Heidelberg",
pages="449--482",
isbn="978-3-540-24777-7",
doi="10.1007/978-3-540-24777-7_15",
url="https://doi.org/10.1007/978-3-540-24777-7_15"
}

@article{PRXQuantum.1.020319,
  title = {Exploring Entanglement and Optimization within the Hamiltonian Variational Ansatz},
  author = {Wiersema, Roeland and Zhou, Cunlu and de Sereville, Yvette and Carrasquilla, Juan Felipe and Kim, Yong Baek and Yuen, Henry},
  journal = {PRX Quantum},
  volume = {1},
  issue = {2},
  pages = {020319},
  numpages = {14},
  year = {2020},
  month = {Dec},
  publisher = {American Physical Society},
  doi = {10.1103/PRXQuantum.1.020319},
  url = {https://link.aps.org/doi/10.1103/PRXQuantum.1.020319}
}

@article{PhysRevA.92.042303,
  title = {Progress towards practical quantum variational algorithms},
  author = {Wecker, Dave and Hastings, Matthew B. and Troyer, Matthias},
  journal = {Phys. Rev. A},
  volume = {92},
  issue = {4},
  pages = {042303},
  numpages = {10},
  year = {2015},
  month = {Oct},
  publisher = {American Physical Society},
  doi = {10.1103/PhysRevA.92.042303},
  url = {https://link.aps.org/doi/10.1103/PhysRevA.92.042303}
}

\end{document}